\documentclass[prl,amsmath,twocolumn,amssymb,aps,superscriptaddress,floatfix,preprintnumbers,10pt,notitlepage,openany]{revtex4-1}
\usepackage{times}
\usepackage{graphicx}
\usepackage{amsmath, braket, amsfonts}
\usepackage{amssymb}
\usepackage{natbib}
\usepackage{sidecap}
\usepackage{bm, color, ulem}
\usepackage{xcolor}
\usepackage{lipsum}
\usepackage{changes}
\usepackage{titlesec}

\titlespacing{\section}
{0pt}{2.5ex plus .2ex minus .2ex}{2.5ex plus .2ex minus .2ex}

\titlespacing{\subsection}
{0pt}{1.5ex plus .2ex minus .2ex}{1.5ex plus .2ex minus .2ex}

\begin{document}

\title{Observation of Interface Piezoelectricity in Superconducting Devices on Silicon}

\author{Haoxin Zhou}
\affiliation{
Department of Electrical Engineering and Computer Sciences, University of California,  Berkeley, Berkeley, California 94720, USA
}
\affiliation{
 Materials Sciences Division, Lawrence Berkeley National Laboratory, Berkeley, California 94720, USA
}
\affiliation{
Department of Physics, University of California, Berkeley, Berkeley, California 94720, USA
}

\author{Eric Li}
\email{Present address: Department of Electrical Engineering and Computer Science, Massachusetts Institute of Technology, Cambridge, MA 02139}
\affiliation{
Department of Electrical Engineering and Computer Sciences, University of California,  Berkeley, Berkeley, California 94720, USA
}

\author{Kadircan Godeneli}
\affiliation{
Department of Electrical Engineering and Computer Sciences, University of California,  Berkeley, Berkeley, California 94720, USA
}
\affiliation{
 Materials Sciences Division, Lawrence Berkeley National Laboratory, Berkeley, California 94720, USA
}

\author{Zi-Huai Zhang}
\affiliation{
Department of Electrical Engineering and Computer Sciences, University of California,  Berkeley, Berkeley, California 94720, USA
}
\affiliation{
 Materials Sciences Division, Lawrence Berkeley National Laboratory, Berkeley, California 94720, USA
}
\affiliation{
Department of Physics, University of California, Berkeley, Berkeley, California 94720, USA
}

\author{Shahin Jahanbani}
\affiliation{
Department of Physics, University of California,  Berkeley, Berkeley, California 94720, USA
}
\affiliation{
 Materials Sciences Division, Lawrence Berkeley National Laboratory, Berkeley, California 94720, USA
}

\author{Kangdi Yu}
\affiliation{
Department of Electrical Engineering and Computer Sciences, University of California,  Berkeley, Berkeley, California 94720, USA
}
\affiliation{
 Materials Sciences Division, Lawrence Berkeley National Laboratory, Berkeley, California 94720, USA
}

\author{Mutasem Odeh}
\affiliation{
Department of Electrical Engineering and Computer Sciences, University of California,  Berkeley, Berkeley, California 94720, USA
}
\affiliation{
 Materials Sciences Division, Lawrence Berkeley National Laboratory, Berkeley, California 94720, USA
}

\author{Shaul Aloni}
\affiliation{
Molecular Foundry, Lawrence Berkeley National Laboratory, Berkeley, California 94720, USA
}

\author{Sin\'ead Griffin}
\affiliation{
Materials Sciences Division, Lawrence Berkeley National Laboratory, Berkeley, California 94720, USA
}
\affiliation{
Molecular Foundry, Lawrence Berkeley National Laboratory, Berkeley, California 94720, USA
}

\author{Alp Sipahigil}
\email{Corresponding author: alp@berkeley.edu}

\affiliation{
Department of Electrical Engineering and Computer Sciences, University of California,  Berkeley, Berkeley, California 94720, USA
}
\affiliation{
 Materials Sciences Division, Lawrence Berkeley National Laboratory, Berkeley, California 94720, USA
}
\affiliation{
Department of Physics, University of California, Berkeley, Berkeley, California 94720, USA
}

\maketitle

{\bf 
The evolution of superconducting quantum processors is driven by the need to reduce errors and scale for fault-tolerant computation~\cite{GoogleQEC2024}.
Reducing physical qubit error rates requires further advances in the microscopic modeling and control of decoherence mechanisms in superconducting qubits.
Piezoelectric interactions contribute to decoherence by mediating energy exchange between microwave photons and acoustic phonons~\cite{jain_acoustic_2023}.
Centrosymmetric materials like silicon and sapphire do not display piezoelectricity and are the preferred substrates for superconducting qubits.
However, the broken centrosymmetry at material interfaces may lead to piezoelectric losses in qubits.
While this loss mechanism was predicted two decades ago~\cite{ioffe_decoherence_2004, georgescu_surface_2019}, interface piezoelectricity has not been experimentally observed in superconducting devices.
Here, we report the observation of interface piezoelectricity at an aluminum-silicon junction and show that it constitutes an important loss channel for superconducting devices.
We fabricate aluminum interdigital surface acoustic wave transducers on silicon and demonstrate piezoelectric transduction from room temperature to millikelvin temperatures.
We find an effective electromechanical coupling factor of  $K^2\approx 2 \times 10^{-5}\%$ comparable to weakly piezoelectric substrates~\cite{yu_acoustic_2017}.
We model the impact of the measured interface piezoelectric response on superconducting qubits and find that the piezoelectric surface loss channel limits qubit quality factors to $Q\sim10^4-10^8$ for designs with different surface participation ratios and electromechanical mode matching.
These results identify electromechanical surface losses as a significant dissipation channel for superconducting qubits, and 
show the need for heterostructure and phononic engineering to minimize errors in next-generation superconducting qubits.
}

The decoherence of superconducting qubits is attributed primarily to material imperfections and interfaces~\cite{siddiqi_engineering_2021}. 
Among these imperfections, decoherence induced by two-level systems -- material imperfections with discrete energy levels -- poses the most significant limitation for state-of-the-art superconducting qubits~\cite{anderson_anomalous_1972, phillips_two_1987}. 
When a superconducting qubit couples to two-level systems, the latter act as electromechanical transducers, causing energy exchange between the microwave photons and acoustic phonons~\cite{ioffe_decoherence_2004, chenu_two_2019}.
However, acoustic phonon radiation is not solely caused by two-level systems. 
Superconducting qubits fabricated on piezoelectric substrates or coupled to piezoelectric transducers can also dissipate through electromechanical transduction and phonon radiation~\cite{jain_acoustic_2023}. 
Superconducting quantum circuits are often fabricated on centrosymmetric substrate materials, such as silicon and sapphire, where piezoelectricity is nominally absent in the bulk. 
However, the defects and interfaces of these materials often result in deviations from their ideal bulk properties~\cite{turiansky_dielectric_2024,georgescu_surface_2019}.
In particular, piezoelectricity can arise at the interfaces even when the bulk materials are not inherently piezoelectric.
Electromechanical transduction based on charge-transfer-induced Coulomb forces has been demonstrated by intentionally breaking centrosymmetry, such as by introducing impurities~\cite{hwang_transduction_2011, kanygin_localized_2019, ransley_a_2007, yang_piezoelectric_2020} or by driving a current~\cite{park_induced_2022} in non-piezoelectric materials. 
However, electromechanical transduction could, in principle, exist without introducing these perturbations and could potentially be present in high-quality superconductor-substrate interfaces used in superconducting qubits. 

\begin{figure*}[t]
\centering
\includegraphics[width=\textwidth]{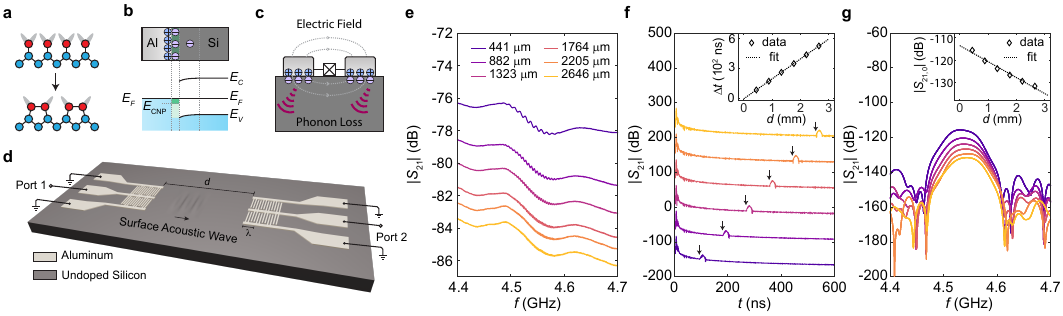}
\caption{\textbf{Observation of interface piezoelectricity via surface acoustic waves.}
\textbf{a}, Schematic of the 1x2 surface reconstruction of silicon (100) surface.
\textbf{b}, Schematic of the energy band diagram at the aluminum-silicon interface and the formation of interface dipoles. The green region indicates the interface electronic states induced by aluminum. $E_C$ and $E_V$ are the energy of the silicon conduction and valence band edges, $E_F$ is the Fermi energy, and $E_{\rm CNP}$ is the charge neutrality point of the silicon interface states.
\textbf{c}, Interface piezoelectricity induced phonon loss in superconducting qubits.
\textbf{d}, Experiment setup. Aluminum IDTs fabricated on silicon transmit and receive surface acoustic waves.
\textbf{e}, Microwave transmission coefficient $|S_{21}|$ as a function of driving frequency measured for devices with different separation distance $d$ on Sample A (Fig.~\ref{fig:S:images}a).
\textbf{f}, Time-domain $|S_{21}|$ as a function of delay time $t$. Each curve is offset by 75~dB for clarity. The black arrows indicate the onset time of surface acoustic wave transmission ($t_{\rm s}$). Inset: $\Delta t = t_{\rm s} - t_{\rm c}$ as a function of $d$. Here $t_s$ ($t_c$) is the onset of transmission mediated by the surface acoustic waves (capacitive crosstalk). Dashed line: Linear fit $d = v \cdot \Delta t$  gives the silicon surface wave velocity $v = 5063$~m/s. 
\textbf{g}, Time-gated $|S_{21}|$ as a function frequency for devices with different $d$. Inset: Time-gated $|S_{21}|$ at the electromechanical resonance as a function of $d$. Dashed line: fit. All measurements are conducted at room temperature. 
}\label{fig:fig1}
\end{figure*}

Here, we characterize the piezoelectric transduction in a widely used heterostructure for superconducting qubit fabrication: the aluminum-silicon junction. 
Crystalline silicon can exhibit complex interface behaviors. 
First, the distinct environments of atoms near the surfaces can induce surface lattice relaxation or reconstruction~\cite{gupta_lattice_1981, lee_reconstruction_1994, kern_elastic_1997, ramstad_theoretical_1995} (Fig.~\ref{fig:fig1}a). 
Theory predicts that lattice relaxation can induce surface piezoelectricity on the sapphire (0001) surface~\cite{shen_a_2010, dai_surface_2011, georgescu_surface_2019}, and a similar effect is expected on silicon surfaces.
Second, the aluminum film alters the electronic structure of silicon near the interface, generating metal-induced gap states~\cite{heine_theory_1965, louie_self_1975, louie_electronic_1976}. 
As undoped silicon has a higher work function than aluminum~\cite{allen_work_1962, crc_handbook_2007}, free electrons transfer from aluminum to silicon so that the heterostructure reaches electrochemical equilibrium while maintaining the continuity of the vacuum energy level across the junction~\cite{grundmann_the_2016} (Fig.~\ref{fig:fig1}b). 
Similar to lattice relaxation, the charge transfer leads to an electric dipole at the interface, which can contribute to a piezoelectric response.
Both mechanisms can induce piezoelectric interface loss in superconducting qubits by resulting in phonon radiation (Fig.~\ref{fig:fig1}c).

\subsection{Piezoelectric Transduction at Aluminum-Silicon Interfaces}

To study the interface piezoelectricity at aluminum-silicon interfaces, we designed and fabricated aluminum interdigital transducers (IDT) with a finger pitch of 1.05~$\mu$m on an undoped silicon (100) substrate. 
The IDTs electromechanically transduce phonons in a surface acoustic wave (SAW) delay-line configuration (Fig.~\ref{fig:fig1}d).
We use a split-finger design~\cite{bristol_applications_1972} to suppress the internal acoustic reflection of the SAW due to mass loading~(Fig.~\ref{fig:S:images}h).
Two mechanisms contribute to the microwave transmission. 
The first is the capacitive crosstalk between the transmitter and receiver IDTs. 
The second, which occurs only in the presence of interface piezoelectricity, is the transduction between microwaves and SAWs. 

Fig.~\ref{fig:fig1}e shows the transmission coefficient measured at room temperature on devices with various separation distances $d$ between the IDTs (Sample A, Fig.~\ref{fig:S:images}a).
We observe oscillations of the transmission coefficient on top of the smooth background around the IDT resonance of $f = 4.55$~GHz.
The oscillation results from the interference between the slow-propagating surface acoustic waves and the capacitive crosstalk~\cite{supp}.
To give a more direct demonstration of the piezoelectric transduction, we convert the data to the time domain by performing a Fourier transform, as shown in Fig.~\ref{fig:fig1}f.
This gives an approximate impulse response of the device.
For all devices, the transmission coefficient peaks at $t = t_{\rm c}=2.5$~ns, which is the microwave propagation delay and is insensitive to $d$.
Remarkably, a second peak appears after a longer delay time $t_{\rm s}$. 
Unlike $t_{\rm c}$, $t_{\rm s}$ is proportional to $d$.
Fitting the data by the relation $d = v \cdot (t_{\rm s} - t_{\rm c})$ gives $v\approx5063$~m/s (Inset of Fig. \ref{fig:fig1}f), close to the 4920~m/s -- the literature reported speed of surface acoustic waves on silicon (100) surface without electrical or mechanical loads~\cite{tarasenko_theoretical_2021}.

To quantitatively study the piezoelectric transduction and SAW propagation, we apply a rectangular filter in the time domain around $t_s$ (or ``time gating'') and transfer it back to the frequency domain.
This procedure removes the contribution from capacitive crosstalk and restores the resonance peak in the frequency-domain response~\cite{yu_acoustic_2017}, as shown in Fig.~\ref{fig:fig1}g.
The transmission coefficient at resonance ($S_{21,0}$) decreases with $d$, which can be attributed to the propagation loss of SAW.
The relation between the $S_{21,0}$ and $d$ can be fit to an exponential function $|S_{21,0}| = Ae^{-d/2l}$ with fitting parameters $A =1.96\times 10^{-6}$ and $l=$~0.6~mm.
The latter corresponds to the decay length of SAW at room temperature.
The exponential behavior suggests that the attenuation of SAW is dominated by the Akhiezer damping theory of elastic waves~\cite{li_attenuation_2016}.

We performed the same measurement on devices fabricated with different methods (See Methods section for details) and observed piezoelectric transduction on all six samples studied~(Fig.~\ref{fig:S:process}).
These findings indicate that interface piezoelectricity is robust against fabrication process variations and the presence of native silicon oxide at the aluminum-silicon interface.
To assess the quality of the aluminum-silicon interface, we used the same substrate for etched IDTs to make superconducting resonators (Fig.~\ref{fig:S:resonator}a and Ref.~\cite{zhang_acceptor_2024}).
The superconducting resonators etched from the same aluminum-silicon heterostructure revealed internal quality factors of $Q_i\approx10^6$ at the single-photon level (Fig.~\ref{fig:S:resonator}b and c). These results show that our observation of interface piezoelectricity is relevant for substrates and fabrication processes used for superconducting quantum devices. 

\begin{figure}[t]
\centering
\includegraphics[width=\columnwidth]{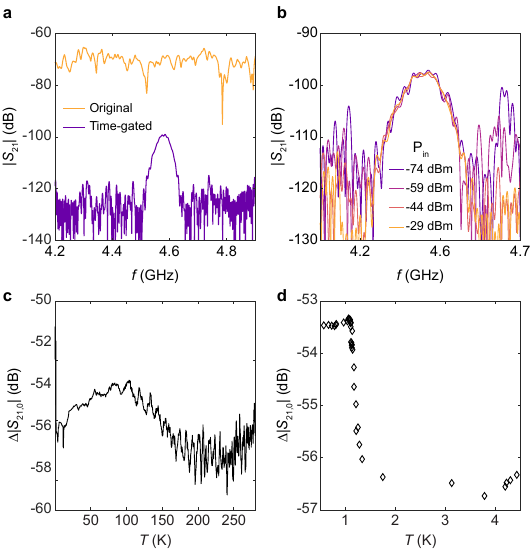}
\caption{\textbf{Interface piezoelectric transduction at cryogenic temperatures.}
\textbf{a}, Original (orange) and time-gated (purple) transmission coefficient in the frequency domain at $T = 30$~mK, measured on Sample B.
\textbf{b}, The time-gated transmission coefficient as a function of frequency $f$ with different excitation powers on Sample B at $T = 20$~mK.
\textbf{c}, The time-gated transmission coefficient ratio between Samples B and Sample D at the electromechanical resonance as a function of temperature. 
\textbf{d}, Same as c, zoom-in on the low-temperature regime. 
}\label{fig:fig2}
\end{figure}

\subsection{Interface Piezoelectricity at Millikelvin Temperatures}
Superconducting qubits operate at cryogenic temperatures, where deviations in electronic and structural properties can influence the piezoelectric response of aluminum-silicon heterostructures.
To study the cryogenic response, we fabricated SAW transducers and packaged them for microwave transmission measurements in a dilution refrigerator  (Sample B, Fig.~\ref{fig:S:images}b).
Fig.~\ref{fig:fig2}a shows the microwave transmission coefficient measured at $T=$~30~mK.
By applying time-gating to remove the contribution of capacitive crosstalk (which is stronger due to the sample geometry and measurement setup), a clear electromechanical resonance peak is resolved, with the resonance transmission coefficient $|S_{21,0}|\approx -99$~dB.
Varying the excitation power by four orders of magnitude does not change $S_{21,0}$ (Fig.~\ref{fig:fig2}b).
The lack of power dependence distinguishes the observed electromechanical transduction from electrostriction and two-level system response, which are nonlinear.
The linear interface piezoelectric surface losses observed here also provide an experimental method to distinguish surface piezoelectric and surface two-level system losses in superconducting devices.  

The measured  $|S_{21}|_\text{max}$ value allows us to use an analytical model~\cite{datta_surface_1986} to calculate the effective electromechanical coupling coefficient $K^2$, the conversion efficiency between electrical and mechanical domains, for the aluminum-silicon IDT (See~\cite{supp} for more details.)
\begin{equation}
K^2 = \frac{1+(2 \pi f_0 C_g Z_0)^2}{2Z_0}\frac{\zeta}{8 \gamma C_g f_0 N L} |S_{21}(f_0)|.
\end{equation}
Here $f_0$ is the resonant frequency. $C_g = 318$~fF is the transducer capacitance. $N =$~50 is the number of finger periods. $Z_0=$~50$~\Omega$ is the reference impedance. $\zeta =$~1.0836 and $\gamma=1.414$ are geometric factors, and $L$ is the propagation loss of SAW.
The Akhiezer damping-induced propagation loss is expected to be negligible at millikelvin temperatures due to the lack of thermal phonons.
We confirmed this by cryogenic distance-dependent transmission coefficient measurements on Sample C (Fig.~\ref{fig:S:images}c), as shown in Fig.~\ref{fig:S:PL}.
We find that $|S_{21,0}|$ is not sensitive to the separation distance $d$, and therefore $L\approx 1$ at low temperatures.
From the results above, we obtain $K^2 \sim 2 \times 10^{-5} \%$ as the effective electromechanical coupling coefficient for aluminum on silicon IDTs.
This value is comparable to weakly piezoelectric substrates such as 4H-SiC~\cite{yu_acoustic_2017}.
The observed electromechanical coupling strength can lead to a significant surface loss channel for superconducting qubits.
We analyze the impact on qubits in a later section.

We study the temperature dependence of the transduction efficiency by comparing the microwave transmission coefficient of the aluminum-on-silicon transducers with a control device (Sample D, Fig.~\ref{fig:S:images}d). This approach allows us to reduce calibration uncertainties due to resistive losses above the superconducting transition of aluminum. 
The control device has the same geometry except that the electromechanical transduction is mediated by a 200-nm piezoelectric aluminum nitride film (Fig.~\ref{fig:S:aln}b).
Since the piezoelectric response of aluminum nitride has a weak temperature dependence~\cite{kazuhiko_temperature_2006}, the ratio of the microwave transmission coefficients at the electromechanical resonance, $\Delta |S_{21,0}| = |S_{21,0}^{\rm Si}|/|S_{21,0}^{\rm AlN}|$ approximates the temperature dependence of the electromechanical transduction at the aluminum-silicon interface.
(See Fig.~\ref{fig:S:aln} for details of the aluminum nitride film and additional data.)
As shown in Fig.~\ref{fig:fig2}c, $\Delta |S_{21}|$ exhibits a non-monotonic dependence on temperature. The variation is less than 6~dB within the experimental temperature range and corresponds to a $<50\%$ change in the effective piezoelectric coefficient. These measurements demonstrate that interface piezoelectricity also has a weak temperature dependence. 
An interesting observation is that $\Delta |S_{21}|$ shows a rapid enhancement at around $T = 1.2$~K, coinciding with the aluminum's superconducting phase transition temperature (Fig.~\ref{fig:fig2}d).
This suggests a potential link between interface piezoelectricity and aluminum's superconducting transition, though the mechanism remains unclear.

\subsection{Microscopic Mechanisms of Interface Piezoelectricity}

\begin{figure}[t]
\centering
\includegraphics[width=\columnwidth]{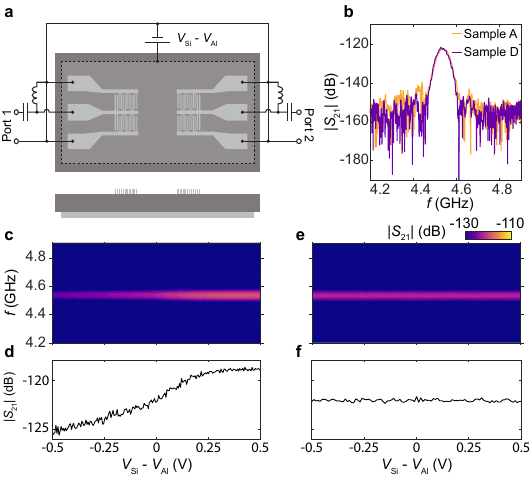}
\caption{
\textbf{Bias field dependence of the interface piezoelectricity.}
\textbf{a}, Schematics of the bias-dependent measurement. Top: measurement wiring. Bottom: cross-section of the device. 
\textbf{b}, Zero-bias time-gated microwave transmission coefficient measured on Samples A and D at room temperature.
\textbf{c}, Time-gated transmission coefficient $|S_{21}|$ as a function of frequency~$f$ and bias voltage $V_{\rm Si} - V_{\rm Al}$ measured on Sample A. The separation distance between the IDTs is $d=1323~\mu$m.
\textbf{d}, Linecut of the data in panel c at the resonant frequency.
\textbf{e}, Same as c, measured on Sample D (Fig.~\ref{fig:S:images}d) where native oxide is removed.
\textbf{f}, Linecut of the data in panel e at the resonant frequency.
}\label{fig:fig3}
\end{figure}

As discussed previously, both the surface lattice relaxation and work function mismatch-induced interlayer charge transfer can induce electromechanical transduction. In the latter case, the charge on the silicon side can either exist as thermally activated space charges or hosted by metal-induced gap states.
The distribution of the thermally activated space charge is sensitive to temperature~\cite{heiman_space_1965}.
Since we observed a weak temperature dependence (Fig.~\ref{fig:fig2}c), the space charge is less likely to be the dominant mechanism.
To gain further insights into the microscopic origins, we study how the transduction depends on a DC bias between the aluminum IDT and silicon substrate, as shown in Fig.~\ref{fig:fig3}a.
We apply a DC bias voltage to the backside of the substrate while keeping the signal and ground electrodes of the IDTs at DC ground.
We investigate two samples with identical geometry, but one has the native oxide removed by hydrofluoric acid (Sample E, Fig.~\ref{fig:S:images}e).
Without the bias voltage, the two samples show nearly identical transduction efficiency (Fig.~\ref{fig:fig3}b).
The lattice-relaxation-induced transduction should be sensitive to the material surface termination~\cite{georgescu_surface_2019}.
In addition, the finite-element analysis discussed later indicates the piezoelectric transduction observed here is significantly stronger than the prediction in Ref.~\cite{georgescu_surface_2019}.
The lack of dependence of the transduction efficiency on the presence of native oxide and the larger transduction efficiency suggests that lattice relaxation is probably not the dominant contributor to interface piezoelectricity.
When a bias voltage is applied, Sample A, with the native oxide present, shows enhanced piezoelectric transduction when $V_{\rm Si}>V_{\rm Al}$ and suppressed transduction when $V_{\rm Si}<V_{\rm Al}$ (Fig.~\ref{fig:fig3}c and d).
This response is consistent with the interface charge transfer picture, as the charge distribution depends on both the work function mismatch and an external electrical potential difference.
Furthermore, we used the finite-element method to simulate the charge distribution near an ideal aluminum-silicon junction in response to a bias voltage (Fig.~\ref{fig:S:semi}).
The resulting polarity aligns with the experimental results.
However, Sample D, with native oxide removed, does not show a bias-dependence (Fig.~\ref{fig:fig3}e and f).
A possible explanation is that the potential drop across the accumulation layer near the interface is too low to alter the interface charge distribution when no oxide is present.

Considering all the results, the interface charge transfer between aluminum and the metal-induced gap states in silicon is likely the primary contributor to interface piezoelectricity, although further investigation is required for a definitive conclusion.

\begin{figure*}[t]
\centering
\includegraphics[width=13.5cm]{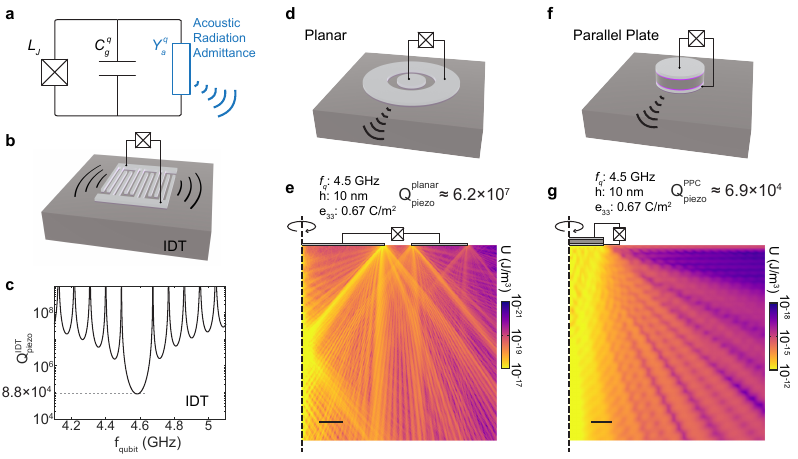}
\caption{
\textbf{Interface piezoelectric surface loss in superconducting qubits.} 
\textbf{a}, A circuit model of the transmon qubit with interface piezoelectric loss. Electromechanical transduction is modeled by an acoustic radiation admittance $Y_a^q(\omega)$. 
\textbf{b}, Schematic of a transmon with an interdigital capacitor. Phonons are radiated as surface acoustic waves near resonance. 
\textbf{c}, The calculated quality factor of a transmon qubit with an interdigital shunt capacitor identical to the IDTs used in the experiment near the electromechanical resonance.
\textbf{d}, A schematic of the transmon qubit with a coplanar shunt capacitor.
\textbf{e}, Simulated spatial distribution of the mechanical energy density within the silicon substrate. The coplanar capacitor is driven at frequency $f = 4.5$~GHz, with the total electrostatic energy equal to that of one microwave photon at $f = 4.5$~GHz. The scale bar represents $50~\mu$m. The positions of the two aluminum electrodes are indicated, though not drawn to scale in the vertical direction.
\textbf{f}, A schematic of the transmon qubit with a parallel-plate shunt capacitor.
\textbf{g}, Same as e, for the parallel-plate capacitor. Scale bar represents $5~\mu$m.
}\label{fig:fig4}
\end{figure*}

\subsection{Impact on Superconducting Qubits}
Despite the uncertainties in the microscopic mechanism, the experimentally observed interface piezoelectricity will constitute a loss channel for superconducting qubits on silicon. 
Here, we adopt the circuit model shown in Fig.~\ref{fig:fig4}a to estimate the impact of interface piezoelectricity on the lifetime of superconducting qubits~\cite{kim_electrical_2008}.
The lossless transmon qubit is characterized by its junction inductance $L_J$ and shunt capacitance $C_g$.
Similar to modeling an antenna with an effective radiation impedance, we use the acoustic radiation admittance $Y_a$ to describe the conversion of the qubit electrical energy into acoustic radiation, i.e., the piezoelectric loss. 
For the IDTs used in the experiment, $Y_a$ can be directly obtained from the effective piezoelectric coupling coefficient $K^2$~\cite{supp}.
We use the experimentally observed $Y_a$ in the IDT geometry to calculate the quality factor of a transmon qubit with a shunt capacitor identical to the IDTs used in the experiment (Fig.~\ref{fig:fig4}b).
The result is shown in Fig.~\ref{fig:fig4}c.
At the electromechanical resonance, the piezoelectric loss-limited quality factor of the qubit is $Q_{\text{piezo}}^{\text{IDT}} \approx 9\times10^4$, indicating a significant loss channel.

Next, we study the interface piezoelectric loss for more commonly used transmon qubit capacitor geometries shown in Fig.~\ref{fig:fig4}d and f. The loss rate induced by interface piezoelectricity in these transmon qubits can differ from the IDT geometry of  Fig.~\ref{fig:fig4}b due to different surface participation and electromechanical mode matching. 
We use finite-element analysis to estimate interface piezoelectric losses in transmon qubits with  125-fF shunting capacitors with concentric coplanar (``planar'', Fig.~\ref{fig:fig4}d) and parallel-plate  (PPC, Fig.~\ref{fig:fig4}f) geometries. 
The axial symmetry is used to simplify the simulation while keeping the footprint comparable to a typical transmon qubit~\cite{krantz_a_2019}.
The dimensions of the two geometries are shown in Fig. \ref{fig:S:qubit_loss}.

Finite-element analysis requires defining a piezoelectric region within the silicon substrate with a fixed thickness and piezoelectric coupling coefficient. 
However, our SAW experiment only determines $K^2$. This effective lumped parameter depends on the distributions of electromagnetic and mechanical fields, as well as the properties and thickness of the piezoelectric material.
The same value of $K^2$ can arise from different combinations of an effective piezoelectric film with thickness $h$ and the piezoelectric coupling coefficient $e_{33}$.

To simulate the qubit loss in different capacitor geometries, we need to know the effective piezoelectric film properties $(h, e_{33})$.
We use finite-element analysis to find combinations of $(h, e_{33})$ that give the experimentally observed $K^2$~\cite{supp} in an IDT (Fig.~\ref{fig:S:model}c and d). 
Next, we use these $(h, e_{33})$  combinations to calculate the qubit loss in different capacitor geometries.
Fig.~\ref{fig:fig4}e and g show the radiated mechanical energy density distribution under the qubit capacitors for $h=10$~nm and $e_{33}=0.67{\rm C/m^2}$ in the planar and PPC geometries.
The interface piezoelectricity-limited quality factors for the planar and parallel plate capacitor are $Q_{\text{piezo}}^{\text{planar}}\approx 6 \times 10^7$ and $Q_{\text{piezo}}^{\text{PPC}}\approx 7 \times 10^4$, respectively. 
The large difference between $Q_{\textrm{piezo}}^{\text{planar}}$ and $Q_{\text{piezo}}^{\text{PPC}}$ arises from their different interface participation ratios.
We repeat the same  $Q_\text{piezo}$ calculations with several different $(h, e_{33})$ combinations that match the experimentally observed  $K^2$ (Fig.~\ref{fig:S:qubit_loss}d). At $f_q = 4.5$~GHz, similar  $Q_\text{piezo}$ values are found in the range of  $Q_{\text{piezo}}^{\text{planar}}\approx 2-6  \times 10^7$ and $Q_{\text{piezo}}^{\text{PPC}}\approx 3-7  \times 10^4$ for the two geometries. We find a small decrease of $Q_\text{piezo}$ for $(h, e_{33})$ combinations with large $h$  and smaller $e_{33}$. This observation is consistent with improved piezoelectric mode matching between the electric and strain fields away from the top surface which is a strain node. 

\subsection{Discussion}

These results establish interface piezoelectricity as a newly identified loss channel for superconducting qubits.
State-of-the-art superconducting qubits on silicon have recently achieved quality factors of $Q\approx 2 \times 10^7$~\cite{tuokkola_methods_2024}.
Our predicted interface piezoelectric surface loss limit of $Q_{\text{piezo}}^{\text{planar}}\approx 6 \times 10^7$ is already of the same order of magnitude.
These suggest that interface piezoelectric surface losses could already be significant in state-of-the-art superconducting qubits.
In addition, we note that the piezoelectric losses do not saturate at high power, which may explain the saturation of high-power quality factors of resonators~\cite{tuokkola_methods_2024}, including our results shown in Fig.~\ref{fig:S:resonator}.

The qubit loss induced by interface piezoelectricity originates from the heterostructure material properties of the qubits rather than material disorders such as defects or amorphous interfaces. As such, even high-quality metal-substrate interfaces will suffer from this loss mechanism. 
Therefore, delicate engineering at the device level is essential to mitigate this loss mechanism. 
These might include shielding qubits from the resonant phonon bath using phononic metamaterials ~\cite{chen_phonon_2023, odeh_non-markovian_2023},  and interface charge transfer engineering by choosing different qubit heterostructures.

On the positive side, the observed reasonably strong and process-insensitive transduction opens up the possibility of realizing electromechanical quantum transduction with the interface piezoelectricity. 
Current approaches for electromechanical transduction require either piezoelectric materials~\cite{mirhosseini_superconducting_2020, meesala_non_2024} or static charge induced by an externally applied voltage~\cite{bozkurt_a_2023, zhao_quantum_2024}, which increase device fabrication and biasing complexity and noise.
Utilizing interface piezoelectricity may eliminate both problems, allowing high-performance electromechanical quantum transducers to be fabricated using a relatively simple and industrially compatible process. Finally, metal-oxide-semiconductor structures could also enhance the charge transfer and transduction strength using large biases for classical transducer and filter applications. 

\section{METHODS}
\subsection{Sample Fabrication}
Sample A, B, C, E, F and G (Fig.~\ref{fig:S:images}) were fabricated on float-zone intrinsic (100) silicon substrates with room temperature resistivity $>10^4~\Omega \cdot$cm.  The aluminum films, unless noted otherwise, are deposited on the (100) surface of an undoped silicon substrate. 
Sample D was fabricated on substrates obtained from Kyma Technologies, which have 200~nm aluminum nitride film on float-zone intrinsic (111) silicon wafers with room temperature resistivity $>10^4~\Omega \cdot$cm. 

Sample A was fabricated by first depositing a 100-nm aluminum film on the backside of the silicon chip using an electron beam evaporator.
The chip was subsequently annealed at 420~$^\circ$C in a forming gas atmosphere to create Ohmic aluminum-silicon contacts.
Finally, the IDTs were formed by depositing a \mbox{50~nm} aluminum film on the front side of the chip with an electron-beam-lithography-defined mask in an electron beam evaporator, followed by a lift-off process.
The chip was exposed to an oxygen plasma before the aluminum deposition to clean the aluminum-silicon interface. 

Sample B and Sample C were fabricated following a two-step lift-off process. 
The IDTs were first formed following the same process for Sample A. 
The coplanar waveguides were then formed by depositing an \mbox{100~nm} aluminum film on the chip with a photolithography-defined mask in an electron beam evaporator, followed by a lift-off process.
The sample was wire-bonded to a printed circuit board for cryogenic measurements.

Sample D was fabricated following the same process as Sample A on the aluminum nitride substrate.

Sample E was fabricated following a lift-off process similar to that for Sample A. 
The difference is that the sample was dipped in a 10:1 buffered hydrofluoric acid solution for 16~s to remove the surface oxide after the oxygen plasma treatment.

Sample F was fabricated using an etch process. 
The substrate was first extensively cleaned with piranha solution, diluted hydrochloric acid solution, and 25:1 hydrofluoric acid solution. 
A 50~nm aluminum film was then deposited using a DC magnetron sputtering system.
Finally, the IDTs were formed by electron beam lithography and a plasma etch.

Sample G was fabricated following a two-step etch process. After the substrate cleaning and aluminum deposition process identical to those for Sample F, the IDTs were formed by electron beam lithography and plasma etch. The coplanar waveguides were formed by photolithography and a wet etch.

The information of Sample A - G is summarized in Table.~\ref{table:S:samples}

The microwave resonator was fabricated with an etch process. After the substrate cleaning and aluminum deposition process identical to those for Sample F, the coplanar waveguide and the microwave resonator were formed by photolithography and a wet etch.

\subsection{Measurement}
Room temperature measurements were performed using a radio-frequency probe station and a vector network analyzer (VNA). 
The microwave power delivered to the end of the probes was 15~dBm.
The bias-dependence measurements were performed using radio-frequency bias tees.
The DC bias voltage was applied using a source meter.
The cryogenic measurements were performed in a dilution refrigerator. 
The power delivered to the wire bonding pads of the devices was \mbox{-29~dBm} unless otherwise indicated. 
A high-electron-mobility transistor (HEMT) is used to amplify the transmitted signal, while circulators are used to isolate the reflection from the HEMT to the device.
The power-dependent quality factor of the microwave resonator was obtained by measuring the transmission coefficient of the device in a dilution refrigerator using a VNA. The details are described in Ref.~\cite{zhang_acceptor_2024}.

\section{Acknowledgements}
We thank Rohin Tangirala for experimental assistance, Francois Leonard for fruitful discussions, as well as Robert Schoelkopf and Srujan Meesala for valuable feedback on the manuscript. This work was primarily funded by the U.S. Department of Energy, Office of Science, Office of Basic Energy Sciences, Materials Sciences and Engineering Division under Contract No. DE-AC02-05-CH11231 in the Phonon Control for Next-Generation Superconducting Systems and Sensors FWP (KCAS23) for device design, modeling, measurements, imaging, and theory. Additional support was provided by the ONR and AFOSR Quantum Phononics MURI program for fabrication process development. The devices used in this work were fabricated at UC Berkeley's Marvell Nanofabrication Laboratory.

\normalem
\let\oldaddcontentsline\addcontentsline
\renewcommand{\addcontentsline}[3]{}
\bibliographystyle{custom}
\bibliography{references}
\let\addcontentsline\oldaddcontentsline

\renewcommand\thefigure{S\arabic{figure}}
\setcounter{figure}{0}
\renewcommand\thetable{S\arabic{table}}
\setcounter{table}{0}

\begin{table*}
\begin{center}
   \caption{Summary of the Samples Studied}
    \begin{tabular}{l|l|l|l|l}
    \label{table:S:samples}
     Sample & Substrate & Fabrication Process & Measurement Environment & Figures\\
     \hline
      A & Silicon & Lift-off & Room temperature & \ref{fig:fig1}(e) - (g), \ref{fig:fig1}(b) - (d), \ref{fig:S:process}(a)\\
      B & Silicon & Lift-off & Cryogenic & \ref{fig:fig2}(a) - (d), \ref{fig:S:aln}(e) - (j)\\
      C & Silicon & Lift-off & Cryogenic & \ref{fig:S:PL}(a) - (d)\\
      D & Aluminum Nitride on Silicon & Lift-off & Cryogenic & \ref{fig:fig2}(c) - (d), \ref{fig:S:aln}(e) - (j)\\
      E & Silicon & Lift-off, Oxide removed & Room temperature & \ref{fig:fig3}(b), \ref{fig:fig3}(e) - (f), \ref{fig:S:process}(b)\\
      F & Silicon & Etch, Oxide removed & Room temperature & \ref{fig:S:process}(c)\\
      G & Silicon & Etch, Oxide removed & Cryogenic & \ref{fig:S:process}(d)\\
    \end{tabular}
\end{center}
\end{table*}

\begin{figure*}[h]
\centering
\includegraphics[width=\textwidth]{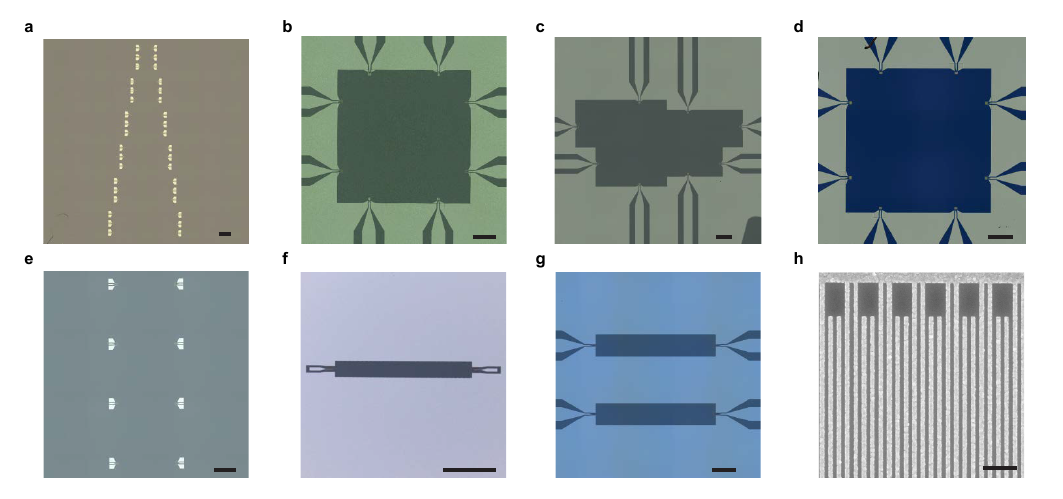}
\caption{\textbf{Optical and scanning electron micrographs of the studied samples.}
\textbf{a} - \textbf{g}, Optical micrographs of Sample A, B, C, D, E, F, and G respectively. The scale bars represent 500~$\mu$m.
\textbf{h}, Scanning electron micrograph of a split-finger transducer. The scale bar represents 1~$\mu$m.
}\label{fig:S:images}
\end{figure*}

\begin{figure*}[h]
\centering
\includegraphics[width=\textwidth]{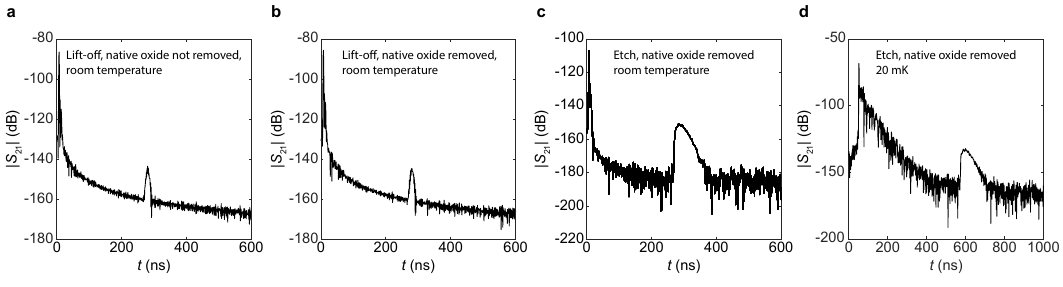}
\caption{\textbf{Interface piezoelectricity of samples fabricated with different processes.}
\textbf{a}, Microwave transmission coefficient $|S_{21}|$ as a function of delay time $t$ measured from Sample A at room temperature. The sample was fabricated following an aluminum lift-off process. The native oxide was not removed before aluminum deposition.
\textbf{b}, Microwave transmission coefficient $|S_{21}|$ as a function of delay time $t$ measured from Sample B at room temperature. The sample was fabricated following an aluminum lift-off process. Native oxide was removed using buffered hydrofluoric acid before aluminum deposition.
\textbf{c}, Microwave transmission coefficient $|S_{21}|$ as a function of delay time $t$ measured from Sample F at room temperature. The sample was fabricated by etching the pre-deposited aluminum film. Due to the limited pattern-transfer resolution, single-finger geometry was used for this sample. Such geometry does not suppress mass-loading-induced reflection of the surface acoustic waves as the transducers with split-finger geometry do; however, this does not affect the conclusion that interface piezoelectricity exists in etched samples.
\textbf{d}, Microwave transmission coefficient $|S_{21}|$ as a function of delay time $t$ measured from Sample G at $T = 20$~mK. The sample was fabricated by etching the pre-deposited aluminum film. 
}\label{fig:S:process}
\end{figure*}

\begin{figure*}[h]
\centering
\includegraphics[width=\textwidth]{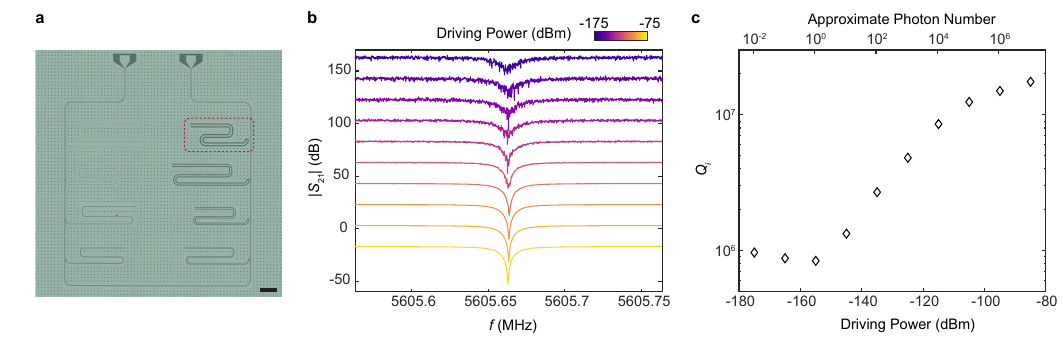}
\caption{
\textbf{Characterizing the quality of aluminum-silicon interface using a coplanar waveguide resonator}
\textbf{a}, Optical micrograph of the sample. The resonator being studied is marked with a red dashed rectangle and has a surface participation ratio of $4.5\times 10^{-4}$~\cite{zhang_acceptor_2024}. The scale bar represents 500~$\mu$m.
\textbf{b}, Microwave transmission coefficient of the device near the resonant frequency at different driving powers.
\textbf{c}, Internal quality factor $Q_i$ as a function of driving power obtained by fitting the data in b. The approximate photon numbers are labeled on the top axis.
}\label{fig:S:resonator}
\end{figure*}

\begin{figure*}[h]
\centering
\includegraphics[width=\textwidth]{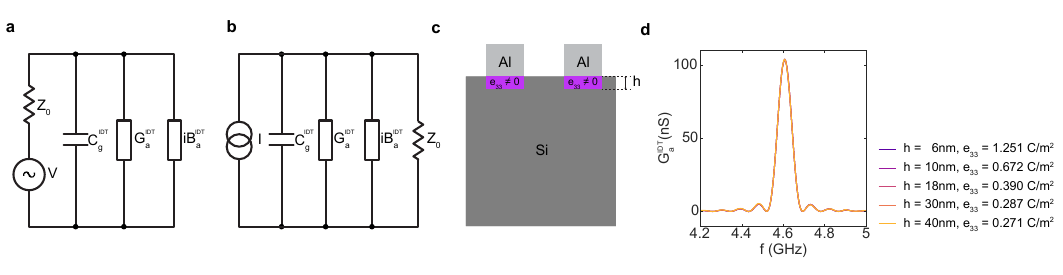}
\caption{\textbf{Circuit models of the IDTs}.
\textbf{a}, Circuit model of the transmitter IDT.
\textbf{b}, circuit model of the receiver IDT.
\textbf{c}, Geometry for searching $(h, e_{33})$ combinations from experimentally determined $K^2$ using finite-element analysis.
\textbf{d}, Simulated acoustic radiation admittance $G_a^\text{IDT}$ of the IDT as a function of frequency. Four $(h, e_{33})$ combinations are found to produce nearly identical $G_a^\text{IDT}(f)$, and therefore, nearly identical $K^2$.
}\label{fig:S:model}
\end{figure*}

\begin{figure*}[h]
\centering
\includegraphics[width=\textwidth]{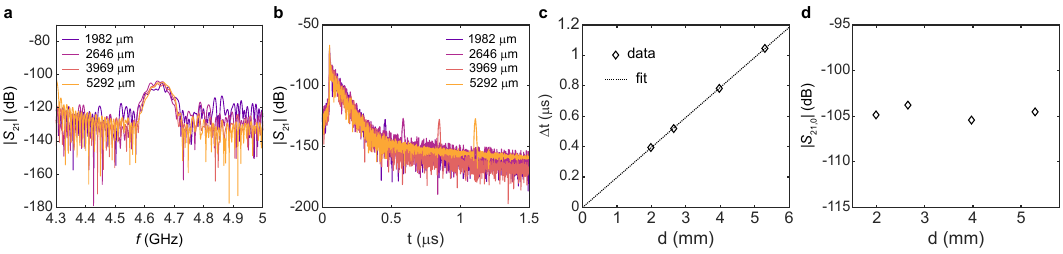}
\caption{
\textbf{Measurement of propagation loss of surface acoustic waves at cryogenic temperature.}
\textbf{a}, Time-gated microwave transmission coefficient near the electromechanical resonance measured from the four delay line IDTs with different separation distance $d$ on Sample C at $T = 20$~mK. No obvious attenuation is observed as $d$ increases.
\textbf{b}, Time-domain microwave transmission coefficient of the same devices in a.
\textbf{c}, Diamonds: $\Delta t = t_{\rm s} - t_{\rm c}$ as a function of $d$. Here $t_s$ ($t_c$) is the onset of transmission mediated by the surface acoustic waves (capacitive crosstalk). Dashed line: Linear fit $d = v \cdot \Delta t$  gives the surface acoustic wave velocity $v = 5065$~m/s. 
\textbf{d}, Microwave transmission coefficient at the electromechanical resonance as a function of $d$.
}\label{fig:S:PL}
\end{figure*}

\begin{figure*}[h]
\centering
\includegraphics[width=\textwidth]{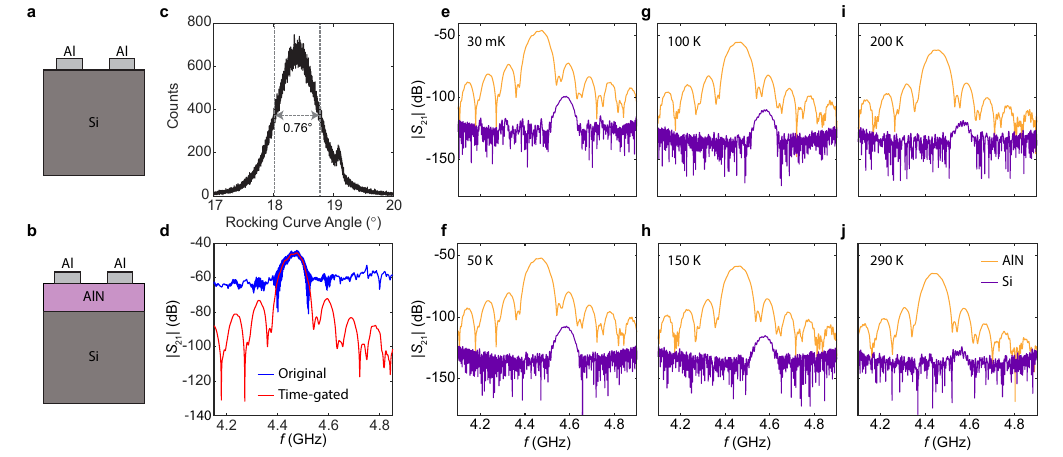}
\caption{
\textbf{Comparison with IDTs fabricated on a piezoelectric aluminum nitride film.}
\textbf{a}, Schematic cross-section of the aluminum-on-silicon surface acoustic wave transducers.
\textbf{b}, Schematic cross-section of the aluminum-on-aluminum nitride surface acoustic wave transducer, used as a reference device (Sample D).
\textbf{c}, X-ray diffraction rocking curve of the AlN film on silicon (111) substrate. The peak corresponds to the AlN (002) plane. The full width at the half maximum is measured to be 0.76$^\circ$, smaller than 1$^\circ$. The result indicates that the 33-component of the piezoelectric coefficient is close to that of single-crystal aluminum nitride~\cite{naik_measurements_2000, jonisch_piezoelectric_2006, mishin_sputtered_2003}.
\textbf{d}, Original and time-gated microwave transmission coefficient of Sample D (aluminum-on-aluminum nitride transducer) measured at $T = 30$~mK.
\textbf{e} - \textbf{j}, Comparison of time-gated microwave transmission coefficient between Sample B (aluminum-on-silicon transducer) and Sample D (aluminum-on-aluminum nitride transducer) at different temperatures indicated in the figure.
}\label{fig:S:aln}
\end{figure*}

\begin{figure*}[h]
\centering
\includegraphics[width=\textwidth]{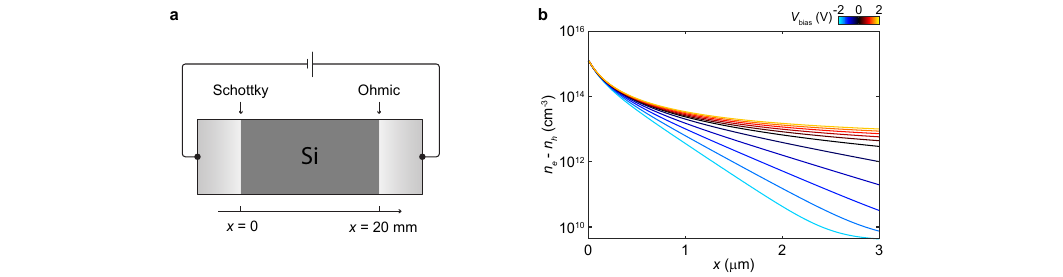}
\caption{
\textbf{Simulated charge distribution in undoped silicon near the aluminum-silicon interface at room temperature using a finite element solver.}
\textbf{a}, Schematic of the system studied. An ideal Ohmic contact is used to set the electrical potential on the other side of the silicon domain. 
\textbf{b}, Excess electron density $n_e - n_h$ as a function of $x$. Different curves correspond to different bias voltages applied across the structure. Here, $n_e$ and $n_h$ are the electron and hole density, respectively.
As expected, an accumulation layer is formed near the interface.
The charge density near the interface increases when a positive voltage is applied on the back electrode and decreases when a negative voltage is applied.
}\label{fig:S:semi}
\end{figure*}

\begin{figure*}[h]
\centering
\includegraphics[width=\textwidth]{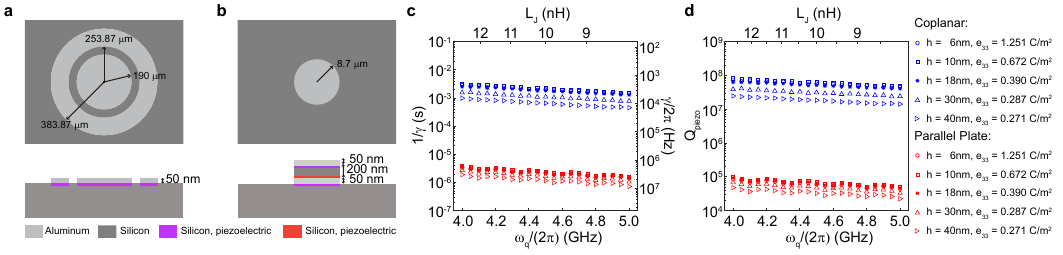}
\caption{
\textbf{Details of interface piezoelectricity induced surface loss in superconducting qubits.} 
\textbf{a}, Top and cross-section view of the axial-symmetric coplanar capacitor. A piezoelectric region is added at the aluminum-silicon interface. The plots are not to scale.
\textbf{b}, Same as panel a, for the axial-symmetric parallel-plate capacitor (PPC). The plots are not to scale. Since the direction of the electrical dipole moment near the interface between the upper aluminum plate and the silicon dielectric layer is opposite to that between the lower aluminum plate and the silicon dielectric layer, the signs of $e_{33}$ for these two interfaces (rendered in purple and red, respectively) are the opposite.
\textbf{c}, Loss rate and interface-piezoelectric-loss-limited relaxation time of the transmon qubits as a function of the qubit frequency. Several different combinations of piezoelectric region thickness $h$ and piezoelectric coupling coefficient $e_{33}$ that produce the same $K^2$ are used for the calculation.
\textbf{c}, Interface piezoelectric loss-limited quality factor $Q_\textrm{piezo}$ as a function of the qubit frequency. 
}\label{fig:S:qubit_loss}
\end{figure*}

\clearpage
\pagebreak
\onecolumngrid
\begin{center}
\textbf{\large Supplementary information for ``Observation of interface piezoelectricity in superconducting devices on silicon'' }\\[5pt]
\begin{quote}
 {\small 
}
\end{quote}
\end{center}
\setcounter{equation}{0}
\setcounter{table}{0}
\setcounter{page}{1}
\setcounter{section}{0}
\makeatletter
\renewcommand{\theequation}{S\arabic{equation}}
\renewcommand{\thefigure}{S\arabic{figure}}
\renewcommand{\thepage}{S\arabic{page}}

\section{Interference between crosstalk and piezoelectricity-induced microwave transmission}

The transmission coefficient $S_{21}$ of the IDTs is defined to be
\begin{equation}
    S_{21}(\omega) = \frac{V_2^-}{V_1^+},
\end{equation}
where $V_1^+$ is the voltage wave propagating into the launching IDT, $V_2^-$ is the voltage wave propagating out of the detecting IDT, and $\omega = 2\pi f$ is the angular frequency.

When a steady drive $V_1 = V_1^+ + V_1^- = |V_1|e^{i(\omega t + \phi_1)}$ is applied to the launching IDT, the forward-propagating voltage of the detecting IDT will have the form
\begin{equation}
    V_2^- = \left( V_e^+(\omega, V_1)e^{i\omega d/v_e}+V_a^+(\omega, V_1)e^{i\omega d/v_a}\right)e^{i\omega t},
    \label{eq1}
\end{equation}
where $V_e^+(V_1, \omega)$ and $V_a^+(V_1, \omega)$ are the amplitudes of the forward-propagating voltage waves induced by electromagnetic crosstalk and interface piezoelectricity, respectively. $v_e$ and $v_a$ are the speeds of electromagnetic waves and surface acoustic waves. The specific form of $V_e^+$ and $V_a^+$ depends on the device geometry and dielectric environment. In addition, $V_a^+$ depends on the piezoelectric coupling strength. In our design, $V_a^+$ is only significant at the electromechanical resonant frequency, while $V_e^+$ is not sensitive to frequency near the electromechanical resonance.

Because $v_e$ and $v_a$ are different, the two terms in Eq. \ref{eq1} interfere constructively or destructively depending on $\omega$
\begin{align}
    |V_2^-|^2 & = \left( V_e^+(\omega, V_1)e^{i\omega d/v_e}+V_a^+(\omega, V_1)e^{i\omega d/v_a}\right) \left( V_e^+(\omega, V_1)e^{-i\omega d/v_e}+V_a^+(\omega, V_1)e^{-i\omega d/v_a}\right) 
\nonumber \\ \label{eq2}
              & = (V_e^+)^2+(V_a^+)^2 + 2 V_e^+ V_a^+ \cos{\left(\omega (d/v_e - d/v_a)\right)},
\end{align}
which generates the oscillating features shown in Fig. \ref{fig:fig1}e. Eq. \ref{eq2} also shows that the oscillating frequency is proportional to $d$, consistent with the experimental results.

\section{Extracting $K^2$ from microwave transmission coefficient}
In this section, we discuss the circuit model of IDTs and the method used to extract the effective coupling coefficient $K^2$ from the microwave transmission coefficient $S_{21}$. 
The circuit model used is based on the classical analysis of IDT~\cite{smith_IDTmodel_1969, datta_surface_1986}.
Here, we give a brief introduction to it.

Fig.~\ref{fig:S:model}a and b show the circuit model of the loaded transmitter and receiver IDTs, respectively.
The model of a single transducer includes the geometric capacitance $C_g$ and a complex admittance $Y_a^\text{IDT} (\omega) = G_a^\text{IDT}(\omega) + iB_a^\text{IDT}(\omega)$ that characterizes the electromechanical transduction. 
$G_a^\text{IDT}$ and $B_a^\text{IDT}$ are the real and imaginary parts of the effective admittance, respectively.
At the electromechanical resonant frequency $f = f_0 = \omega_0 / 2 \pi$, $B_a=0$, and~\cite{hines_a_1993}
\begin{equation}
    G_a^\text{IDT}(\omega_0)
    = 8K^2\gamma C_g^\text{IDT} f_0 N/ \zeta. \label{ga}
\end{equation}
Here, $N$ is the number of the IDT periodic units.
$\gamma$ and $\zeta$ are dimensionless parameters determined by the device geometry. 
For the split-finger transducers, $\gamma = 1.0836$ and $\zeta = 1.414$~\cite{hines_a_1993}.

During the measurement, the transmitter is connected to a voltage source with impedance $Z_0 = 50 ~\Omega$.
The energy consumed by $G_a^\text{IDT}$ is converted to mechanical energy of the surface acoustic waves.
The receiver is driven by an effective voltage-controlled current source and is loaded by $50 ~\Omega$. The energy consumed by the load represents the electrical energy converted from the mechanical power source.
The insertion loss of the transmitter is
\begin{equation}
    {\rm IL}_t = \frac{2G_a^\text{IDT} Z_0}{(1+G_a^\text{IDT} Z_0)^2 + [Z_0 (\omega_0 C_g^\text{IDT}+B_a^\text{IDT})]^2},
\end{equation}
which is defined as the ratio of the power converted into the power of the forward-propagating SAW to the maximal available power.
The insertion loss of the receiver ${\rm IL}_r$ is the same as that of the transmitter by reciprocity.
Therefore, the full microwave transmission coefficient is given by
\begin{equation}
    |S_{21}(\omega_0)|
    = ({\rm IL}_t)^{1/2} \cdot L \cdot ({\rm IL}_r)^{1/2} 
    = \frac{2G_a^\text{IDT} Z_0 L}{(1+G_a^\text{IDT} Z_0)^2 + [Z_0 (\omega_0 C_g^\text{IDT}+B_a^\text{IDT})]^2},
    \label{s2g}
\end{equation}
where $L$ is the amplitude propagation loss during SAW propagation.
Since $G_a^\text{IDT} \ll 1/Z_0$, and $B_a^\text{IDT} = 0$ at $f = f_0$,
\begin{equation}
    |S_{21}(\omega_0)| 
    \approx \frac{2 G_a^\text{IDT}(\omega_0) Z_0 L}{1+(\omega_0 C_g Z_0)^2}, \label{s21}
\end{equation}
From Eq.~\ref{ga} and \ref{s21}, we obtain
\begin{equation}
    K^2 = \frac{1+(\omega_0 C_g^\text{IDT} Z_0)^2}{2Z_0}\frac{2 \pi \zeta}{8 \gamma C_g^\text{IDT} \omega_0 N L} |S_{21}(\omega_0)| . \label{k2}
\end{equation}
Eq.~\ref{k2} offers a way to calculate $K^2$ from the measured $S_{21}(\omega_0)$.

The above model ignores the capacitive crosstalk between the transmitter and receiver, which cannot be omitted when piezoelectricity is weak.
Here, we employ the time-gating method to eliminate the contribution of capacitive crosstalk~\cite{yu_acoustic_2017}.
We begin by performing an inverse Fourier transform to convert the data into the time domain.
Next, a bandpass filter is applied in the time domain to filter out the crosstalk contribution, which occurs much earlier than the SAW signal.
Finally, we transform the filtered data back to the frequency domain.
After removing the crosstalk contribution, the spectrum shows a clear maximum at the resonant frequency $f_0 = 4.583$~GHz. As shown in Fig.~\ref{fig:fig2}a, $|S_{21}(\omega_0)| = 1.1221\times 10^{-5}$ or $-99$~dB.

The pass loss is not sensitive to temperature and is extracted from the distance-dependent measurement in Fig.~\ref{fig:fig1}h.

From Fig.~\ref{fig:S:PL}, the propagation loss at cryogenic temperature is negligibie.
Therefore, $L=1$.
Substituting the calculated $L$ into Eq.~\ref{k2} with $N = 50$ and $C_g = 318$~fF, we obtained $K^2 = 2.32 \times 10^{-7}$.

\section{Qubit loss rate induced by the interface piezoelectricity}

In this section, we derive the expression for the loss rate $\gamma$ from the circuit model shown in Fig. \ref{fig:fig4}a.
It contains the circuit diagram of a transmon qubit with the interface-piezoelectricity-induced loss included. 
As discussed in the previous section, the piezoelectric effect can be represented by a parasitic radiation admittance $Y_a^q (\omega) = G_a^q (\omega) + i B_a^q (\omega)$, where $G_a^q$ and $B_a^q$ are the radiation conductance and susceptance, respectively. The total admittance between Node 1 and Node 2 shown in Fig. \ref{fig:fig4}a is
\begin{equation}
\begin{split}
    Y_{12}^q (\omega)
    & = \frac{1}{i \omega L_J} + i \omega C_g^q + G_a^q(\omega) + i B_a^q(\omega)
\\
    & = \frac{1 - \omega^2 L_J [C_g^q + B_a^q(\omega) / \omega] + i \omega L_J G_a^q(\omega)}{i \omega L_J}.
\end{split}
\end{equation}
Since the piezoelectricity is weak, the radiation susceptance $B_a$ is negligible compared to $\omega C_g^q$, i.e.,
\begin{equation}
    Y_{12}^q (\omega) 
    \approx \frac{1 - \omega^2 L_J C_g^q + i \omega L_J G_a^q(\omega)}{i \omega L_J}.
    \label{z_res}
\end{equation}

Eq. \ref{z_res} is the impedance of a parallel resonant circuit with resonant (plasmon) frequency 
\begin{equation}
    \frac{\omega}{2\pi} 
    = \frac{1}{2\pi\sqrt{L_J C_g^q}}
\end{equation}
and quality factor 
\begin{equation}
    Q(\omega) = \frac{\omega C_g^q}{G_{a}^q(\omega)} = \frac{\omega}{\gamma (\omega)},
\end{equation}
where $\gamma(\omega) = G_a^q(\omega) / C_g^q$ is the piezoelectric loss rate.
The corresponding relaxation time of the resonator is given by
\begin{equation}
    T_1 (\omega_q) = \frac{1}{\gamma(\omega_q)} = \frac{C_g^q}{G_a^q(\omega_q)}\label{decay_rate},
\end{equation}
where $\omega_q = \omega_p - e^2 / 2 C_g$ is transmon qubit frequency.
Therefore, if $G_a^q (\omega_q)$ is known, $Q(\omega_q)$ or $T_1(\omega_q)$ can be obtained.

For the superconducting qubit with a shunt capacitor that has the same geometry as the IDTs studied in the experiment, $G_a^q$ at the electromechanical resonance ($\omega = \omega_0$) can be obtained from $S_{21}$ (Eq.~\ref{s2g}).
At other frequencies~\cite{datta_surface_1986},
\begin{equation}
    G_a^q(\omega) = G_a^q(\omega_0)\frac{[N \pi (\omega-\omega_0) / \omega_0]^2}{\sin^2[N \pi (\omega-\omega_0) /\omega_0]}.
\end{equation}

For the superconducting qubits with different (i.e. not IDT) geometries, $G_a^q(\omega)$ can be obtained numerically using the finite-element method.
The first step is calculating $G_a^q(\omega)$ of an aluminum-on-silicon IDT with the same geometry used in the experiment.
The piezoelectric coupling is simulated by assigning a finite $e_{33}$ of a thin silicon layer with thickness $h$ under the aluminum pattern that matches the measured $G_a^q(\omega_0)$.
Different combinations of $h$ and $e_{33}$ can produce the same $G_a^q(\omega_0)$.Fig.~\ref{fig:S:model}d shows the four combinations we use for the qubit loss simulation.
Then, we use finite-element multiphysics simulations to calculate the admittance of the desired qubit shunt capacitor with a piezoelectric silicon layer near the aluminum-silicon interface that has thickness $h$ and piezoelectric coupling coefficient $e_{33}$ using the values obtained.
These simulations give the $G_a^q(\omega)$ of the qubit shunt capacitor from which the qubit quality factor is calculated.

\end{document}